\documentclass{jpp}
\usepackage{graphicx}
\usepackage{epstopdf, epsfig}
\usepackage[bookmarks=false]{hyperref}

\usepackage{xcolor}
\usepackage{bm}
\usepackage{amsmath}
\usepackage{natbib}
\usepackage{color}

\newcommand{\dd}{\mathrm d}
\graphicspath{{./figures/}}
\usepackage{physics}
\usepackage{amsmath}
\usepackage{enumitem}
\newcommand\ut{\mathsfbi{U}}
\newcommand\iit{\mathsfbi{I}}

\newcommand{\pder}[2][]{\frac{\partial#1}{\partial#2}}

\graphicspath{{./figures/}}

\newcommand\thefont{\expandafter\string\the\font}

\newcommand{\vpp}{v_{\perp}}
\newcommand{\vp}{v_{\parallel}}
\newcommand{\pp}{_{\perp}}
\newcommand{\pa}{_{\parallel}}

\newcommand{\uvec}[1]{\hat{\rm{\mathbf{#1}}}}
\newcommand{\divv}{\nabla\cdot}

\newcommand{\divvv}{\nabla_{v}\cdot}
\newcommand{\graddv}{\nabla_{v}}

\shorttitle{Cyclotron cooling in electron-ion plasmas}
\shortauthor{D. Kennedy and P. Helander}

\title{Coulomb collisions in strongly anisotropic plasmas \\ I. Cyclotron cooling in electron-ion plasmas }

\author{D. Kennedy\aff{1}\corresp{\email{daniel.kennedy@ipp.mpg.de}}
	\and P. Helander\aff{1}}

\affiliation{\aff{1}Max Planck Institute for Plasma Physics, D-17491 Greifswald, Germany}

\begin{document}

\maketitle

\begin{abstract}
	The behaviour of a collisional plasma which is optically thin to cyclotron radiation is considered, and the distribution functions accessible to it on the various timescales in the system are calculated. Particular attention is paid to the limit in which the collision time exceeds the radiation emission time, making the electron distribution function strongly anisotropic. Unusually for plasma physics, the collision operator can nevertheless be calculated analytically although the plasma is far from Maxwellian. The rate of radiation emission is calculated and found to be governed by the collision frequency multiplied by a factor that only depends logarithmically on plasma parameters.
\end{abstract}

\section{Introduction}

In the presence of a strong magnetic field, charged particles execute helical gyromotion around the magnetic field lines. The Lorentz force acts on particles in a direction perpendicular to both the magnetic field and the particles' motion through them, causing them to accelerate, and in turn to release energy in the form of electromagnetic waves, ``cyclotron radiation''. Plasmas with sufficiently large particle density, such as fusion plasmas, are optically thick to such emissions, meaning that any energy released through this process is simply re-absorbed back into the plasma. However, certain plasma systems are optically thin to cyclotron emission and this radiated energy can therefore be lost to the surroundings. These optically thin plasmas are the focus of this current work. It is the purpose of the present paper to show how these emissions lead to the manifestation of anisotropy in the plasma distribution function, and how the distribution evolves in these regimes. 

In this work, paper (I) of this series, our contributions are: (1) to show that cyclotron emission results in strongly anisotropic distribution functions on the radiation timescale; (2) to calculate the evolution of the distribution function under collisional scattering which, in the absence of any radiation terms, acts to drive the plasma towards a Maxwellian; (3) to show that this behaviour manifests itself under very general conditions; and (4) to present the applications and the limitations of this theory. In the companion work, paper (II), we apply this theory to the first laboratory electron-positron plasma experiment. We begin in Section \ref{sec:the_collisionless_system_in_brief} by briefly recapitulating the single particle picture of an electron emitting cyclotron radiation, before using this description to explore the relaxation of the distribution function in a collisionless plasma.  In Section \ref{sec:lowest_order_collisional_kinetic_equation}, collisions are introduced back into the picture and used to show that an initially isotropic distribution will relax to a distribution which is Maxwellian over the perpendicular velocity. In computing this distribution we introduce a free function $T_{\perp}$; in Sections \ref{sec:evolution_of_the_perpendicular_density} and \ref{sec:limiting_forms} the evolution of this free function is explored. In Section \ref{sec:general_magnetic_geometry}, we show that the conclusions ascertained in the previous sections actually hold in quite general magnetic geometry; subsequently, we discuss some of the broad classes of applications in Section \ref{sec:applicability_of_this_theory}. 

\section{The collisionless system in brief}
\label{sec:the_collisionless_system_in_brief}

Whilst we endeavour to provide a physical understanding of this theory in general magnetic geometries, let us first understand the basic premise before embarking on a more detailed calculation. To this end, we begin with a straight, constant magnetic field and, for the time being, neglect collisions in the plasma. 

The fundamental aim of this section is thus to understand the following: (a) In the most simple case, cyclotron emission is responsible for an exponential decay of the perpendicular kinetic energy on the radiation timescale; and (b) the inclusion of cyclotron emission, and the absence of collisions, results in a strongly anisotropic distribution function. 

\subsection{Cyclotron cooling}

The emission of cyclotron radiation by particles in the plasma gives rise to a reaction force which must be included in the kinetic equation. The theory of relativistic plasmas, accounting for this radiation reaction by including the Abraham–Lorentz reaction force in the kinetic equation has been developed by \citet{Andersson2001}, \citet{Hazeltine_Mahajan}, and subsequent authors. Here, we begin by specialising these results to a non-relativistic plasma.

The non-relativistic Abraham-Lorentz reaction force is given by 
\begin{equation}
\bm{K} = \frac{e^{2}}{6\pi\epsilon_{0}c^{3}}\dot{\bm{a}}
\end{equation}
where $\bm{a}$ is the acceleration vector and $e$ is the charge. Following the arguments of the aforementioned authors, it can be seen that, to leading order in a small gyroradius expansion, the change in perpendicular energy, $w_{\perp},$ of a non-relativistic point charge as it accelerates in a magnetic field is given by Larmor's formula
\begin{equation}
\left(\frac{\dd w_{\perp}}{\dd t}\right)_{\text{rad}} = - \frac{e^{2}a^{2}}{6\pi \epsilon_{0}c^{3}}, \quad w_{\perp} = \frac{mv_{\perp}^{2}}{2}, 
\end{equation}
where $\bm{v}$ is the particle velocity vector and $m$ is the rest mass. Throughout this work, the subscripts $\perp$ and $\parallel$ indicate directions perpendicular and parallel to the magnetic field, respectively. The centripetal acceleration of a particle undergoing gyromotion is given by
\begin{equation}
a = \frac{eB}{m}v_{\perp},
\end{equation}
and hence it follows that
\begin{equation}
\left(\frac{\dd w_{\perp}}{\dd t}\right)_{\text{rad}} = - \frac{e^{4}B^{2}}{3\pi\epsilon_{0}(mc)^{3}} w_{\perp}.
\label{cooling_equation}
\end{equation}
Simply put, the plasma will radiate its perpendicular kinetic energy on a timescale given by the radiation time 
\begin{equation}
\tau_{r} = \frac{3\pi\epsilon_{0}(mc)^{3}}{e^{4}B^{2}}.
\end{equation} 

It is perhaps easiest to understand the influence of radiative emission when the change of energy is rewritten in terms of the perpendicular and parallel components of the velocity. We thus obtain 
\begin{equation}
\left(\frac{\dd \vpp}{\dd t}\right)_{\text{rad}} = -\frac{\vpp}{2\tau_{r}}, \quad \left(\frac{\dd \vp}{\dd t}\right)_{\text{rad}} = 0.
\label{reaction_force}
\end{equation} 
Equations (\ref{reaction_force}) summarise the average effect of radiation reaction on the plasma. Fundamentally, it is these equations which will give rise to anisotropy in the distribution function. 

It is pertinent to point out at this stage, that collisions, whose effect is to isotropise the plasma, can mediate this cooling process due to the scattering of the velocity vector; as a result, allowing the conversion of energy between perpendicular and parallel components. We will return to this discussion in Section \ref{sec:lowest_order_collisional_kinetic_equation}, following a discourse on the simpler collisionless case.

\subsection{Collisionless kinetic equation}

The kinetic equation governing the evolution of the distribution function $f(\bm{r},\bm{v},t)$ in the absence of collisions is given by
\begin{equation}
\pder[f]{t} + \divv (\bm{v}f) + \divvv \left[\frac{q}{m}(\bm{E} + \bm{v}\cross\bm{B})f + \left(\frac{\dd \bm{v}}{\dd t}\right)_{\text{rad}}f\right] = 0
\label{2.5}
\end{equation}
where $\graddv$ stands for the velocity derivative 
\begin{equation}
\graddv = \uvec{x}\pder{v_{x}} + \uvec{y}\pder{v_{y}} + \uvec{z}\pder{v_{z}}.
\end{equation}
Cyclotron emission is included in the kinetic equation through the inclusion of (\ref{reaction_force}) in the usual collisionless Vlasov equation. 

We will immediately specialise to the case where the magnetic field is constant, $\bm{B} = B\uvec{b}$, and there is zero electric field, $\bm{E} = \bm{0}.$ Noting that, in this limit, we have assumed the plasma is homogenous and there is no spatial dependence of the distribution function. We then write equation (\ref{2.5}) in cylindrical $\bm{v}-$space coordinates $(\vpp,\alpha,\vp)$ to obtain
\begin{equation}
\pder[f]{t} + \frac{1}{\vpp}\pder{\vpp}(\vpp\dot{v}_{\pp}f) + \pder{\alpha} (\dot{\alpha}f)  = 0,
\label{kinetic_equation_straight} 
\end{equation}
where we have used that $\dot{{v}}_{\parallel} = 0.$ The final term in this equation can then be eliminated by averaging over the gyroangle $\alpha.$

It is fruitful to write the resulting equation in terms of the perpendicular energy of the particles, so to obtain
\begin{equation}
\pder[f]{t}- \frac{w_{\perp}}{\tau_{r}}\pder[f]{w_{\perp}} = \frac{f}{\tau_{r}}.
\end{equation}
This equation can then be easily solved via the method of characteristics; the general solution is given by
\begin{equation}
f(w_{\perp},w_{\parallel},t) = F_{0}\left(w_{\perp}\mathrm{e}^{t/\tau_{r}},w_{\parallel}\right) \mathrm{e}^{t/\tau_{r}}.
\end{equation}
That is, the distribution function will be a function of the parallel and perpendicular kinetic energies, the latter of which will decay exponentially quickly on the radiation timescale. 

As an illustrative example, if the initial distribution is Maxwellian, then 
\begin{equation}
f(w_{\perp},w_{\parallel},t) = n \left(\frac{m}{2\pi T}\right)^{3/2} \exp \left[ - \frac{1}{T} \left(w_{\perp}\mathrm{e}^{t/\tau_{r}} + w_{\parallel}\right) + \frac{t}{\tau_{r}}\right],
\end{equation}
which corresponds to a bi-Maxwellian
\begin{equation}
f(w_{\perp},w_{\parallel},t) = n \left(\frac{m}{2\pi T_{\perp}^{2/3}T_{\parallel}^{1/3}}\right)^{3/2} \exp \left(-\frac{w_{\perp}}{T_{\perp}} - \frac{w_{\parallel}}{T_{\parallel}}\right)
\end{equation}
with $T_{\parallel} = T = \text{constant}$ and 
\begin{equation}
T_{\perp}(t) = T \mathrm{e}^{-t/\tau_{r}}.
\end{equation}

 At this stage of our investigation, we have been able to accomplish the first fundamental aim of this paper; namely, we now understand the following. Firstly, in the most simple case, cyclotron emission is responsible for an exponential decay of the perpendicular energy on the radiation timescale whilst the parallel energy is kept constant. Secondly, the inclusion of radiative emission, and the absence of collisions, results in a strongly anisotropic distribution function on the radiation timescale. 
 
\subsection{Validity of the collisionless approach}
\label{subsec:validity}
A caveat here is that, of course, many plasmas are in fact collisional. It is therefore expected that eventually collisions will come into play and must be taken into account. 

Following \citet{Braginskii1965}, the conventional definition of the electron-ion and electron-electron collision times are given by
\begin{equation}
\tau_{ei} = \frac{6\sqrt{2}\pi^{3/2}\epsilon_{0}^{2}{m_{e}^{1/2}}T_{e}^{3/2}}{ Ze^{4} n_{e} \ln \Lambda}, \quad \tau_{ee} = Z\tau_{ei}
\end{equation}
respectively, 
where  $\ln \Lambda$ is the Coulomb logarithm. We will also assume bulk neutrality throughout this work so that $n_{e} = Zn_{i}.$ From the above expressions, we can see that the collision time will decrease as the plasma cools. 

We might therefore envision the following scenario. A plasma might begin in a regime where the radiation time $\tau_{r}$ is smaller than the initial collision time $\tau_{c} = \text{min}(\tau_{ee},\tau_{ei})$ and hence the collisionless theory will be applicable at first. However, the collision time itself will decrease as the plasma cools and hence this assumption will be violated after sufficient time has elapsed. As a result, collisional effects must be taken into consideration. 

We thus imminently turn our attention to the collisional problem. However, before doing so, let us allow ourselves a small digression to say something about the nature of cyclotron radiation.

 \subsection{Cyclotron cooling and plasma waves}

Before we embark on solving the collisional kinetic equation, let us first turn our attention to the nature of the electromagnetic waves in play. The cyclotron radiation which is being generated has to propagate through our optically thin plasma as some electromagnetic plasma wave \citep{Krall1986}.

For the reasons outlined in the previous section, we declare an interest in optically thin plasmas where the radiation time is (at least initially) much shorter than the collision time. That is, we declare an interest in the limit 
\begin{equation}
	\frac{\tau_{r}}{\tau_{c}} = \frac{mn\ln \Lambda}{\pi \epsilon_{0}B^{2}}\left(\frac{c}{v_{\text{th}}}\right)^{3} = \frac{\ln \Lambda}{\pi} \left(\frac{\omega_{p}}{\Omega_{c}}\right)^{2}\left(\frac{c}{v_{\text{th}}}\right)^{3} \ll 1,
\end{equation}
where have introduced the two characteristic frequencies in the plasma; the plasma frequency and the cyclotron frequency 
\begin{equation}
	\omega_{p}^{2} = \frac{ne^{2}}{m\epsilon_{0}}, \quad
	\Omega_{c} = \frac{eB}{m}.
\end{equation}
Thus, it is easy to see from this ordering that electromagnetic waves with frequencies comparable to the cyclotron frequency are almost light waves:
\begin{equation}
	\omega^{2} (\sim \Omega_{c}^{2}) = \omega_{p}^{2} + k^{2}c^{2} \implies \omega \simeq kc.
\end{equation}
That is, our cyclotron radiation is simply photon emission which can immediately escape the plasma by liberty of the optical thinness. 

We now turn our attention to the collisional problem.

\section{Lowest order collisional kinetic equation}
\label{sec:lowest_order_collisional_kinetic_equation} 

Our aim now, is to deduce the class of distribution functions to which a radiating collisional plasma can relax. When collisions are retained, the kinetic equation satisfied by the distribution function $f$ in straight field lines is 
\begin{equation}
\pder[f]{t} - \frac{1}{\tau_{r}} \pder{\mu}\left(\mu f\right) =  {C}(f), \quad \mu = \frac{mv^{2}}{2B},
\label{KE}
\end{equation}
where we have opted to write this equation in terms of the magnetic moment $\mu$ instead of the perpendicular velocity. The distribution function can now be changed both via radiation (the second term on the left-hand-side) and also via collisions (the term on the right-hand-side). The Landau collision operator for Coulomb interaction \citep{LDLandau1936} is ${C} = {C}_{ee} + {C}_{ei}$ with 
\begin{equation}
C_{ab}(f) = \sigma_{ab} \nabla \cdot \int ff^{\prime} \mathsfbi{U} \cdot \left(\graddv \ln f - \graddv^{\prime}\ln f^{\prime}\right) \, \dd^{3}\bm{v}^{\prime},
\end{equation}
and
\begin{equation}
\sigma_{ee} = \sigma = \frac{n_{e}e^{4}\ln \Lambda}{8\pi\epsilon_{0}^{2}m_{e}^{2}}, \quad \sigma_{ei} = Z\sigma,
\end{equation}
if $f$ is normalised so that 
\begin{equation}
\int f \, \dd^{3}\bm{v} = 1.
\end{equation}
Here we have also introduced $\mathsfbi{U},$ the second-rank tensor
\begin{equation}
\ut(\bm{u}) = \frac{u^{2} \iit - \bm{u}\bm{u}}{u^{3}},
\end{equation}
where $\bm{u} = \bm{v} - \bm{v}^{\prime}$ is the difference in velocity vectors between colliding particles, and $\iit$ is the identity matrix.

The collision frequency is of order $\sigma/v_{\text{th}a}^{3}$ and we are interested in the limit $\sigma\tau_{r}/v_{\text{th}a}^{3} \ll 1,$ where $v_{\text{th}a}$ is the thermal velocity of species $a$. That is, for the reasons outlined in Section \ref{sec:the_collisionless_system_in_brief}, we are declaring an interest in optically thin plasmas where the radiation time is (at least initially) much shorter than the collision time. The key idea here is that, in this limit, by the time collisions enter the picture our plasma will have already radiated a large fraction of its perpendicular energy through the cyclotron cooling process outlined earlier. In this limit $f$ will be strongly anisotropic, due to the cyclotron cooling process, and we write 
\begin{equation}
\bm{v}_{\perp} = \epsilon\bm{x}_{\perp}, \quad \bm{u} = \bm{v}_{\pa} - \bm{v}_{\pa}^{\prime} + \epsilon (\bm{x}_{\perp} - \bm{x}_{\perp}^{\prime}), \quad \epsilon \ll 1.
\end{equation} 
Our strategy is then to expand the collision operators in terms of this small parameter. 

{It is important to point out here that $\epsilon,$ effectively a measure of $v_{\perp}/\vp,$ might itself be expected to change in time, which could have ramifications later in our analysis. However, these concerns can be mollified by noting that the window of time in which this theory is valid is only a few collision times. Thus we expect $\epsilon$ will not change appreciably and will certainly remain $\epsilon \ll 1.$ This is explored further in Section \ref{subsec:validity_collisional}. In what follows, $\epsilon$ is treated as a constant.}

We begin with the electron-electron collision operator. To lowest order in $\epsilon$ we obtain 
\begin{equation}
C_{ee}(f) \simeq \frac{\sigma}{\vpp}\pder{\vpp}\bm{v}_{\pp}\cdot \graddv f \int \frac{f^{\prime}}{u} \, \dd^{3} \bm{v}^{\prime}.
\label{3.7}
\end{equation}
Great care must be taken when dealing with the integral in equation (\ref{3.7}) which is divergent when $\vp = \vp^{\prime}.$  We may evaluate
\begin{equation}
\int \frac{f^{\prime}}{u} \, \dd^{3}\bm{v} = \int \frac{f^{\prime} \, \dd^{3}\bm{v}}{\sqrt{(\vp - \vp^{\prime})^{2} + \epsilon^{2}(\bm{x} - \bm{x}^{\prime})^{2}}} = 2 \int_{-\infty}^{\infty} f^{\prime}(\vp,\vpp^{\prime}) | \ln \epsilon| 2\pi \vpp^{\prime}\,\dd\vpp^{\prime} \equiv g(\vp)
\end{equation}
since, by partial integration, for any suitably well-behaved $f$ we have
\begin{equation}
\int_{0}^{\infty} \frac{f(x)}{\sqrt{x^{2} + \epsilon^{2}}} \, \dd x = \left[f(x)\ln (x + \sqrt{x^{2} + \epsilon^{2}})\right]_{0}^{\infty} - \int_{0}^{\infty} \frac{\dd f}{\dd x} \ln (x + \sqrt{x^{2} + \epsilon^{2}}) \, \dd x
\end{equation}
where \begin{equation}
\int_{0}^{\infty} \frac{\dd f}{\dd x} \ln (x + \sqrt{x^{2} + \epsilon^{2}}) \, \dd x 
\end{equation}
remains finite as $\epsilon \rightarrow 0.$ 
Thus, it follows that 
\begin{equation} 
C_{ee}(f) \simeq \frac{\sigma g(\vp)}{\vpp} \pder{\vpp} \vpp \pder[f]{\vpp}.
\end{equation}

The electron-electron collision term is larger than the electron-ion collision term, 
\begin{align}
C_{ei}(f) &= Z \sigma \pder{\bm{v}}\cdot \int \frac{v^{2}\mathsfbi{I} -\bm{v}\bm{v}}{v^{3}}ff^{\prime}\left(\graddv \ln f - \graddv^{\prime} \ln f^{\prime} \right) \, \dd^{3}\bm{v}^{\prime}, \\
&= Z\sigma \divvv \left(\frac{v^{2}\mathsfbi{I}-\bm{v}\bm{v}}{v^{3}} \cdot \graddv f\right),\\
&= Z\frac{\sigma}{\vpp}\pder{\vpp}\left(\frac{\bm{v}_{\pp}\cdot \graddv f}{\vp}\right),
\end{align}
provided $Z = O(1).$ This can be clearly seen by noting that
\begin{equation}
	\frac{C_{ei}}{C_{ee}} \sim Z \epsilon.
\end{equation}

The lowest-order kinetic equation thus becomes 
\begin{equation}
-\frac{1}{\vpp}\pder{\vpp}\left(\frac{\vpp^{2}}{2}f\right) = \frac{\tau_{r}\sigma g(\vp)}{\vpp} \pder{\vpp} \vpp \pder[f]{\vpp},
\end{equation}
which can be integrated to give
\begin{equation}
f(\vp,\vpp,t) = C(\vp,t) \exp \left(-\frac{m\vpp^{2}}{2T_{\perp}(\vp,t)}\right), \quad T_{\pp} = 2\sigma \tau_{r} m g(\vp,t). 
\end{equation}
The integration constant $C$ is determined by the requirement that
\begin{equation}
N(\vp,t) = \int_{0}^{\infty} f 2\pi\vpp \, \dd \vpp = \frac{2\pi T_{\pp}}{m}C = 4\pi \sigma \tau_{r}g(\vp) C 
\label{Ndef}
\end{equation}
should equal $g(\vp)/2|\ln \epsilon|,$ which gives $C = 1/(8\pi\sigma\tau_{r}|\ln\epsilon|)$ and thus 
\begin{equation}
f(\vp,\vpp,t) = \frac{1}{8\pi\sigma\tau_{r}|\ln \epsilon|}\exp\left(-\frac{m\vpp^{2}}{2T_{\pp}(\vp,t)}\right).
\label{distributuion}
\end{equation}
It is interesting to note that the distribution of particles over perpendicular velocities is Maxwellian, but not for the usual reason. Normally, this happens because of a balance between two terms in the collision operator, describing friction (drag) and energy diffusion, respectively. The former term slows particles down and the latter increases their average energy. The two terms balance exactly for a Maxwellian. In (\ref{distributuion}), radiative energy loss has replaced the collisional friction, and the result is a Maxwellian in $\vpp$ with a different (lower) perpendicular temperature than in the purely collisional case. 

We note that the distribution function only depends on $\vp$ through the function $T_{\perp}(\vp,t).$ We can determine this function by taking a moment of the kinetic equation.

\section{Evolution of the perpendicular density}
\label{sec:evolution_of_the_perpendicular_density}
The moment $N,$ defined in equation (\ref{Ndef}) can be seen to satisfy 
\begin{equation}
\pder[N]{t} = \int_{0}^{\infty} C(f) \, 2\pi\vpp \, \dd \vpp.
\label{a}
\end{equation} {In solving this equation, we obtain the temperature $T_{\perp}$ through (\ref{Ndef}) and can thus ascertain the distribution function $f$ through use of (\ref{distributuion}).} Let us first remark on the nature of (\ref{a}). In deriving (\ref{a}), we have integrated the kinetic equation (\ref{KE}) over perpendicular velocities. In doing so, we have eliminated the radiation term and instead isolated the slow evolution of the distribution over parallel velocities. We can think of the evolution of an initial isotropic distribution function, in the collisional regime, as being partitioned into two disparate stages: 
\begin{enumerate}[topsep=10pt,itemsep=1ex,partopsep=1ex,parsep=1ex]
	\item \enskip Firstly, the distribution function will quickly radiate most of its perpendicular energy and relax to a form which is Maxwellian over perpendicular velocity as per (\ref{distributuion}).
	\item \enskip Secondly, on a longer timescale, the distribution over parallel velocities will evolve according to (\ref{a}). 
\end{enumerate}
It is this slow behaviour (ii) that we are now interested in probing. We will soon see that the precise nature of this evolution depends on the specific ordering satisfied by the plasma parameters.  

In order to ascertain the time evolution of $N,$ we thus require the following moments of the collision operators:
\begin{equation}
I_{ee} \equiv \int C_{ee}(f) \, \dd^{2} \bm{v}_{\perp}, \quad I_{ei} \equiv \int C_{ei}(f) \, \dd^{2} \bm{v}_{\perp}.
\end{equation}
It can be shown that the leading order contributions to these two terms are given by
\begin{align}
	I_{ee} = -\sigma |\ln \epsilon| \pder[^{2}N^{2}]{\vp^{2}}, \quad I_{ei} =Z\sigma \pder{\vp}\left(\frac{N}{\vp|\vp|}\right). \label{Imoments}
\end{align}
The details of the calculation are relegated to Appendix \ref{app1}. It then follows that (\ref{a}) becomes
\begin{equation}
\pder[N]{t} = -\sigma|\ln\epsilon| \pder[^{2}N^{2}]{\vp^{2}} + 2Z\sigma\pder{\vp}\left(\frac{N}{\vp|\vp|}\right)
\end{equation} or, more compactly, we can write
\begin{equation}
\pder[N]{\tau} = \pder{\vp}\left(\frac{N}{\vp|\vp|} - \alpha \pder[N^{2}]{\vp}\right), \quad \tau = \frac{3}{4}\sqrt{\pi} v_{\text{th}}^{3} \frac{t}{\tau_{ei}}, \quad \alpha = |\ln \epsilon| \frac{1}{2Z}.
\label{parabolicPDE}
\end{equation}
Once again, we have been able to successfully reduce our integro-differential evolution equation (\ref{a}) into an equation involving only differential operators (\ref{parabolicPDE}). The two terms on the right-hand side of (\ref{parabolicPDE}) have very different character. The first one describes friction on the electrons from collisions with the ions, which causes the former to slow down in the parallel direction. As usual for Coulomb collisions, the collision frequency decreases with increasing speed, which here causes the singularity at $\vp = 0.$ The second term describes electron-electron collisions, whose effect is to scatter the velocity in a peculiar way, causing ``anti-diffusion'' in $\vp.$ This phenomenon can be understood as follows. Most electron-electron collisions occur between particles with similar velocities $\vp \simeq \vp^{\prime},$ and the local (in parallel velocity space) collision frequency is thus proportional to $N,$ making this term quadratic in $N$ (particles with given $ \vp$ mostly collide with each other). The effect of such collisions is to convert parallel kinetic energy to perpendicular energy, and thus to increase $T_{\pp},$ which according to (\ref{Ndef}) leads to an increase in $N.$ A local accumulation of particles somewhere in parallel velocity space will thus tend to grow at the expense of neighbouring regions, making the distribution function undergo ``anti-diffusion'' in $\vp.$ 

Despite its relatively simple form, (\ref{parabolicPDE}) is a non-linear parabolic partial differential equation. It seems impossible to make analytical progress in full generality here but nevertheless some insight into the system can be gleaned by appealing to various limiting forms. 

\section{Limiting forms of the perpendicular density} 
\label{sec:limiting_forms}
We note that, from (\ref{Imoments}), it is easy to see that the moments of the collision operator (which gives rise to the different terms in (\ref{parabolicPDE})) are ordered
\begin{equation}
	\frac{I_{ei}}{I_{ee}} \sim \frac{Z}{|\ln \epsilon|}. 
\end{equation}
This is to be compared and contrasted to the ordering of the collision operators themselves: \begin{equation}
	\frac{C_{ei}}{C_{ee}} \sim Z \epsilon.
\end{equation}
Based on these different orderings, we can turn to solving equation (\ref{parabolicPDE}) in two distinct limits:
	\begin{alignat}{2}
	&(A): \quad\quad |\ln \epsilon| \ll Z \ll \frac{1}{\epsilon}, \\
		&(B): \quad\quad Z \ll |\ln \epsilon| \ll \frac{1}{\epsilon}. 
	\end{alignat}
In both of these limits, we are justified in dropping the electron-ion collision operator at lowest order as discussed in Section \ref{sec:lowest_order_collisional_kinetic_equation}. This approximation gave rise to the rapid evolution of the distribution function towards (\ref{distributuion}) at lowest order. 

\subsection{(A): neglecting the contribution from electron-electron collisions}

In limit (A), we are justified we can neglect the contribution of $I_{ee}$ whilst retaining that of $I_{ei},$ noting that this is still consistent with the derivation of (\ref{distributuion}) providing the required ordering is satisfied. In this instance, (\ref{parabolicPDE}) becomes
\begin{equation}
\pder[N]{\tau} = \pder{\vp}\left(\frac{N}{\vp^{2}}\right).
\end{equation}
The equation has been reduced to a quasilinear PDE and so is amenable to solution by the method of characteristics. 
We thus obtain the general solution 
\begin{equation}
N(\vp,\tau) = \vp^{2}F\left(\frac{\vp^{3}}{3}  + \tau\right).
\label{5.2}
\end{equation}
So, if we start from a Maxwellian distribution initially 
\begin{equation}
N(\vp,0) = \left(\frac{m}{2\pi T_{\pa}}\right)^{1/2}\exp\left(-\frac{m\vp^{2}}{2T_{\parallel}}\right),
\end{equation}
then 
\begin{equation}
N(\vp,\tau) = \left(\frac{m}{2\pi T_{\pa}}\right)^{1/2}  \frac{\vp^{2}}{(\vp^{3} + 3\tau)^{2/3}}  \exp\left(-\frac{m}{2T_{\parallel}}(\vp^{3} + 3\tau)^{2/3}\right).
\end{equation}

From this equation, we can also obtain an expression for the perpendicular temperature 
\begin{equation}
T_{\perp}(\vp,\tau) = 4\sigma \tau_{r}m|\ln \epsilon|\left(\frac{m}{2\pi T_{\pa}}\right)^{1/2}  \frac{\vp^{2}}{(\vp^{3} + 3\tau)^{2/3}}  \exp\left(-\frac{m}{2T_{\parallel}}(\vp^{3} + 3\tau)^{2/3}\right).
\end{equation}
It is important to recall that this solution was for $\vp > 0,$ which means that, despite appearances, both $T_{\perp}$ and $N$ are continuous. For $\vp < 0 $ we obtain the same solution as above with $\vp \rightarrow -\vp.$

\subsubsection{Rate of energy loss}

A quantity of direct experimental interest is the amount of power radiated by the plasma. This can be calculated by looking at the rate of change of energy loss
\begin{equation}
P = -\frac{\dd W}{\dd t},
\end{equation}
where we have defined the total thermal energy stored in the plasma (per unit volume):
	\begin{equation}
	W = \int \frac{1}{2} mnv^{2}  f \, \dd^{3}\bm{v}.
	\end{equation}
The energy stored can be seen to satisfy the equation	\begin{equation}
	\frac{\dd W}{\dd t} = -\frac{W_{\perp}}{\tau_{r}}, \quad W_{\perp} = \int \frac{1}{2}mnv_{\perp}^{2} f \, \dd^{3}\bm{v}.
	\label{508}
	\end{equation}
Thus, it suffices to calculate $W_{\perp}$ in order to find the power radiated by the plasma. This is given by evaluating
\begin{align} 
W_{\perp} &= \int \frac{1}{2}mn\vpp^{2} f \, \dd^{3} \bm{v}, \nonumber \\ &= \int \frac{mn}{8\sigma\tau_{r}|\ln\epsilon|} \vpp^{3} \exp \left(-\frac{m\vpp^{2}}{2T_{\pp}(\vp,t)}\right) \, \dd \vp \dd \vpp, \nonumber \\ 
&= \int \frac{nT_{\pp}(\vp,t)^{2}}{4\sigma\tau_{r}m|\ln\epsilon|} \, \dd\vp,
\end{align}
where we have used equation (\ref{distributuion}) to carry out the $\vpp$ integral.
We arrive at
	\begin{equation}
	\frac{W_{\perp}(y)}{W_{\perp}(0)} = \frac{2}{\sqrt{\pi}}\int_{0}^{\infty} \frac{x^{4}}{(x^{3}+y)^{4/3}}\exp \left(-(x^{3} + y)^{2/3}\right) \, \dd x,
	\label{512}
	\end{equation}
	where we have normalised $\vp$ and $\tau$ to the electron thermal velocity by writing \begin{equation}
	\vp = v_{\text{th}}x, \quad 3\tau = v_{\text{th}}^{3}y, \quad v_{\text{th}} = \sqrt{\frac{2T_{\pa}}{m}}.
	\end{equation} The integral (\ref{512}) cannot be expressed in terms of standard mathematical functions. Instead, we must turn to numerically evaluating this function. Firstly however, we can determine the behaviour of (\ref{512}) in the long-time and short-time limits. 

In the limit $y \gg 1$ we obtain (see Appendix \ref{appB_long}):
\begin{equation}
	\frac{W_{\pp}(y)}{W_{\pp}(0)} \simeq \left(\frac{2}{3\pi}\right)^{8/9} \Gamma \left(\frac{5}{3}\right) \left(\frac{t}{\tau_{ei}}\right)^{-7/9} \exp\left( -\left(\frac{9\sqrt{\pi}}{4}\frac{t}{\tau_{ei}}\right)^{2/3}\right).
	\label{515_text}
\end{equation}

In the limit $y \ll 1$ we obtain (see Appendix \ref{appB_short}):
\begin{equation}
	\frac{W_{\pp}(t)}{W_{\pp}(0)} \simeq 1 - \frac{\Gamma(5/3) \Gamma(2/3)}{\Gamma(7/3)} \left(\frac{128}{3\pi}\right)^{1/6} \left(\frac{t}{\tau_{ei}}\right)^{1/3}. \label{522_text}
\end{equation}

\begin{figure}
	\begin{center}
	\includegraphics{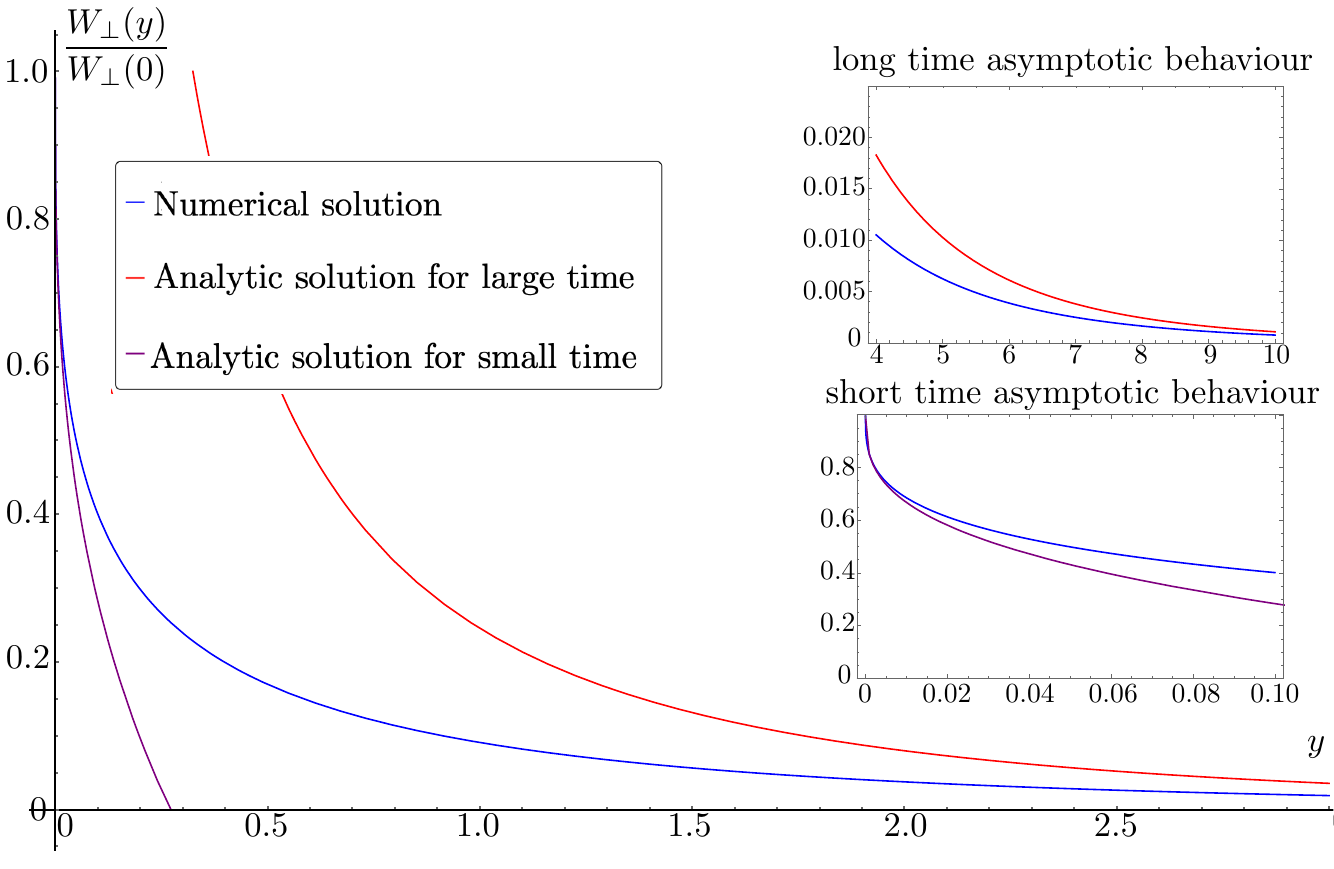}
	\end{center}
	\caption{The ratio of perpendicular energy at normalised time $y = (9\sqrt{\pi}/4)t/\tau_{ei}$ to the initial perpendicular energy, a quantity which is proportional to the rate of energy loss in the plasma. The blue curve is the calculated rate of energy loss from equation (\ref{512}). The red curve is the asymptotic solution for $y \gg 1$ given by equation (\ref{515_text}). The purple curve is the asymptotic solution for $y \ll 1$ given by equation (\ref{522_text}). This figure is for the perpendicular energy loss in limit (A).}
	\label{figure1}
\end{figure}

In figure (\ref{figure1}) we show the full numerical solution of equation (\ref{512}) as well as the analytic solution for the long-term and short-term behaviour, given by (\ref{515_text}) and (\ref{522_text}) respectively. Some caution must be taken in interpreting this figure, and one must remember that the solution of equation (\ref{512}) is only valid on timescales longer than the radiation time. Note that the quantity plotted on the vertical axis, $W_{\perp}(y)/W_{\perp}(0),$ is directly proportional to the emitted power as can be seen from equation (\ref{508}). The abscissa is simply a scaled time coordinate: $y = (9\sqrt{\pi}/4)t/\tau_{ei}.$

Although at a first glance it may seem as though the energy loss rate would be faster than exponential, this is of course not the case when the solution is restricted to the region where the orderings are valid.

\subsection{(B): neglecting the contribution from  electron-ion collisions}

In limit (B), when $n_{i} Z^{2} / n_{e} \ll | \ln \epsilon |$, we are justified in neglecting the contribution from $I_{ei}$ and obtain a relatively simple equation:
\begin{equation}
\pder[N]{\tau} = -\alpha \pder[^{2} N^{2}]{\vp^{2}}.
\label{illposed}
\end{equation}
This is a non-linear parabolic PDE which essentially describes backwards diffusion in $\vp$ space. Indeed, this is simply the backwards heat equation in a medium where the diffusion coefficient varies linearly with temperature. As such, this problem is ill-posed. 

{A simple solution exists (given suitable initial data) but the general solution will not depend continuously on the auxiliary data as it describes a reverse diffusion process meaning that arbitrarily small perturbations in the initial conditions will be amplified and can lead to even infinitesimal-time singularities. This means that it is possible to prepare arbitrarily small perturbations to the initial conditions (likely in any given Sobolev norm), which are sufficiently wiggly in higher derivatives to ensure the solution blows up before any given time $t > 0.$} 

In order to find this singular solution, we consider the problem
\begin{equation}
\pder[N]{\tau} - \beta \pder{\vp}\left(N\pder[N]{\vp}\right) = 0, \quad \beta = - 2\alpha
\label{parabolic_PDE}
\end{equation}
with 
\begin{equation}
\int_{-\infty}^{\infty} N(\vp, \tau) \, \dd \vp = 1.
\label{parabolic_BC}
\end{equation}

Equation (\ref{parabolic_PDE}) admits a similarity solution. To see this, we introduce a dilation transformation 
\begin{equation}
z = \delta^{a}\vp, \quad s = \delta^{b}\tau, \quad v(s,z) = \delta^{c}N (\delta^{-a}z,\delta^{-b}s).  
\end{equation}
Our problem then becomes 
\begin{equation}
\delta^{b-a} \pder[v]{s} - \beta \delta^{2a} \pder[^{2}]{z^{2}}\left(\frac{1}{2}\delta^{-2c}v^{2}\right) = 0.
\end{equation}
From this, we see that equation (\ref{parabolic_PDE}) is invariant under the dilation transformation provided $b -c = 2(a-c)$, i.e. $c = 2a - b.$ 

We therefore pose the self-similar ansatz 
\begin{equation}
N(\vp,\tau) = \tau^{(2a-b)/c}y(\xi), \quad \xi = \frac{\vp}{\tau^{a/b}}.
\end{equation} 
The condition (\ref{parabolic_BC}) then gives \begin{equation}
\int_{-\infty}^{\infty} N(\vp,\tau) \, \dd \vp = 1 \implies \tau^{3(a/b)-1} \int_{-\infty}^{\infty} y(\xi) \, \dd \xi = 1.
\end{equation}
This condition must hold for all $\tau > 0$ and hence we have $3a=b.$ So we can write 
\begin{equation}
N(\vp,\tau) = \tau^{-1/3}y(\xi), \quad \xi = \frac{\vp}{\tau^{1/3}}.
\end{equation} 
It is of course no coincidence to have arrived at this particular self-similar ansatz. We could have also posited this solution simply by noting that the quantity   $\xi = v_{\parallel}^{3}/\tau$ is dimensionless, which we could have deduced from equation (\ref{5.2}). 

The derivatives then transform as 
\begin{equation}
\pder[N]{\tau} = -\frac{1}{3}\tau^{-4/3} (y + \xi y^{\prime}),
\end{equation}
\begin{equation}
\pder[^{2}]{\vp^{2}}\left(\frac{1}{2}N^{2}\right) = \frac{1}{2}\tau^{-2/3} (2yy^{\prime\prime} + 2 y^{\prime2} )\tau^{-2/3}
\end{equation}
where the prime notation now denotes differentiation with respect to $\xi.$ 

Equation (\ref{parabolic_PDE}) becomes 
\begin{equation}
\beta y y^{\prime\prime} + \beta y^{\prime 2} + \frac{1}{3}y + \frac{1}{3}\xi y^{\prime} = 0,
\end{equation}
or equivalently 
\begin{equation}
3\beta (yy^{\prime})^{\prime} + (\xi y)^{\prime} = 0.
\end{equation}
Direct integration gives
\begin{equation}
3\beta yy^{\prime} + \xi y = k_{0}.
\end{equation}
{The integration constant $k_{0}$ is chosen such that the flux of particles $N \frac{\dd N}{\dd \vp}$ is continuous at the edge of the support where $N$ goes to zero. In our case, this is the same constant that one gets from imposing zero flux at the symmetry point because the flux has to vanish when either $N=0$ (at the edge) or when $\frac{\dd N}{\dd \vp} = 0$ (at a symmetry point).}

This equation can then be integrated once again
\begin{equation}
y(\xi) = -\frac{1}{6\beta}\xi^{2} + k_{1}
\end{equation} 

In order to satisfy the requirement that $N \rightarrow 0$ as $| \vp | \rightarrow \infty,$ we write the solution as
\begin{equation}
N(\vp,\tau) = \begin{cases}
\frac{\tau^{-1/3}}{6\beta}\left(A^{2} - \left(\frac{\vp}{\tau^{1/3}}\right)^{2}\right), \quad &|\vp| < A\tau^{1/3} \\ 0 \quad &|\vp| > A\tau^{1/3},
\end{cases}
\end{equation} 
where $A = \sqrt[3]{9\beta/2}$ so as to satisfy (\ref{parabolic_BC}).

We can thus obtain a weak, in the sense that the first derivative is discontinuous, solution of equation (\ref{parabolicPDE}) in the limit where electron-ion collisions are neglected. It is imperative to mention that this solution holds only when $\tau < 0.$ In essence, what we have done in this section is to solve equation an ill-posed backwards diffusion equation (\ref{illposed}) by solving instead a well-posed forward diffusion equation (\ref{parabolic_PDE}) and running the solution backwards in time. The rationale in adopting this approach is to motivate the correct boundary conditions to apply to the equation.

As this problem is ill-posed, our self-similar solution requires very specific initial conditions. An arbitrarily small perturbation (in any given Sobolev norm) can cause the solution to fail and develop an infinitesimal-time singularity. Of course the self-similar solution is still of value, and the formation of singularities is somewhat fictitious since our assumption that $\nabla_{\pa} \ll \nabla_{\pp}$ will certainly break down before any singularity arises. The solution obtained is of the correct general form (i.e., it attains a maximum for $\vp = 0,$ and satisfies $N \rightarrow 0$ as $|\vp|\rightarrow 0$) as to possibly be approached from more general initial conditions. Such solutions are not uncommon in plasma physics and the generation of sharp structures by resistive or ambipolar diffusion has also been investigated \citep{Low}.

It is also fitting to remark here that including higher powers of the expansion parameter $\epsilon$ in our expansion of the collision operator (i.e., in the calculation of $I_{ee}$) would usually bring higher order derivatives of $N$ into equation \ref{illposed} which would regularize the partial differential equation. This is a possible avenue of further exploration with the caveat that the underlying equations would then likely be rendered too difficult to solve with analytical techniques.

As before, one can calculate the perpendicular temperature 
\begin{equation}
T_{\perp}(\vp,\tau) = \begin{cases}
2\sigma\tau_{r}m|\ln\epsilon|\frac{\tau^{-1/3}}{3\beta}\left(A^{2} - \left(\frac{\vp}{\tau^{1/3}}\right)^{2}\right), \quad &|\vp| < A\tau^{1/3} \\ 0 \quad &|\vp| > A\tau^{1/3}.
\end{cases}
\end{equation} 

\subsubsection{Rate of energy loss}

Following the previous subsection, one can calculate the rate at which energy is radiated from the plasma by first calculating $W_{\perp}$ and then appealing to equation (\ref{508}). This is given by 
\begin{equation}
W_{\perp} = \int \frac{nT_{\pp}(\vp,t)^{2}}{4\sigma\tau_{r}m|\ln \epsilon|} \, \dd\vp = 4 mn \sigma \tau_{r} |\ln \epsilon| \left(\frac{3}{4\tau A^{3}}\right)^{2} \int_{-\infty}^{\infty} (A^{2} \tau^{2/3} - \vp^{2})^{2}\, \dd\vp
\end{equation} 
which can be evaluated exactly to give 
\begin{equation}
W_{\perp} = 8 m n\sigma \tau_{r}|\ln \epsilon| \left(\frac{3}{4\tau A^{3}}\right)^{2} \int_{0}^{A\tau^{1/3}}(A^{2}\tau^{2/3} - \vp^{2}) \, \dd\vp = \frac{12}{5} \frac{mn \sigma \tau_{r} |\ln \epsilon|}{A\tau^{1/3}}.
\end{equation}
This leads to an energy loss rate of
\begin{equation}
-\frac{\dd W}{\dd t} = \frac{12}{5} \left(\frac{1}{2Z}\right)^{1/3} \frac{mn\sigma^{2/3}|\ln \epsilon|}{At^{1/3}},
\end{equation}
which can be written in terms of the Braginskii electron-electron collision time $\tau_{ee}$ as 
\begin{equation}
-\frac{\dd W}{\dd t} =  - \frac{6}{5}\pi^{1/3} n_{e}T_{\parallel e} |\ln \epsilon|^{2/3} \left(\frac{1}{t\tau_{ee}^{2}}\right)^{1/3}.
\label{eepower}
\end{equation}
At first glance, this equation might appear somewhat peculiar. We must first of course recall that our solution was valid only when time is run backwards, and hence the $t$ appearing on the right-hand-side is negative; as a result, the total energy does indeed decrease as expected.

\subsection{Validity of the collisional approach; a self-accelerating process}
\label{subsec:validity_collisional}

We have found the solution to (\ref{parabolicPDE}) in two different limits based on the plasma parameters; thus elucidating the plasma distribution function in the collisional regime (\ref{distributuion}). Throughout this work, a number of orderings have been invoked and we thus turn our attention to the validity of our solution; specifically, when it is valid to treat $\epsilon$ as constant. Thus far we have focussed on the evolution of the plasma on timescales $t$ with 
\begin{equation}
	0 < \tau_{r} < t \sim \tau_{c}.
\end{equation} In this regime, we treat $\epsilon$ as constant seeing as though any variation is very small. However, as per our earlier remarks in Section \ref{subsec:validity}, we must take care to recall that $\tau_c$ is not constant and will itself decrease as the plasma cools via cyclotron emission. Our entire process (radiative cooling and collisional scattering) is therefore self-accelerating. Collisional scattering converts parallel kinetic energy to perpendicular kinetic energy. This perpendicular energy is then radiated, lowering the temperature of the plasma and in turn increasing the collisional scattering rate. As a result, the plasma will only remain in this regime for a few collision times, before entering a regime where 
\begin{equation}
\tau_{c} < \tau_{r} < t.
\end{equation}
In this final regime, any remaining parallel kinetic energy will be converted to perpendicular kinetic energy and then radiated. The total energy of the plasma will thus decay exponentially quickly on the radiation timescale.

In essence, we can partition the evolution of the plasma energy into three distinct regimes:\begin{enumerate}[label=(\Roman*.),topsep=10pt,itemsep=1ex,partopsep=1ex,parsep=1ex]
 		\item \enskip Initially, $0 < \tau_{r} < t < \tau_{c}.$ In this regime the behaviour is collisionless as described in Section \ref{sec:the_collisionless_system_in_brief}. The perpendicular energy decays exponentially quickly on the radiation timescale.  
 		\item \enskip  After some time, $\tau_{r} < t \sim \tau_{c} $ and collisional scattering becomes important. This regime is what we have studied in Sections \ref{sec:lowest_order_collisional_kinetic_equation} - \ref{sec:limiting_forms}. This is a self-accelerating process and the plasma will only remain in this regime for a few collision times. 
 		\item \enskip Eventually, $\tau_{c} < \tau_{r} < t.$ In this regime, the distribution function is isotropic (Maxwellian), and any remaining parallel kinetic energy will be converted to perpendicular kinetic energy via collisional scattering and then radiated. 
 	\end{enumerate}

We have now accomplished our second aim; namely, we have been able to calculate the evolution of the distribution function under collisional scattering. Unusually for plasma physics, this was possible even though the distribution function is far from Maxwellian. Thus far, we have made the restrictive assumption of straight field line geometry, we will now ask ourselves whether we can relax this assumption.

\section{General magnetic geometry}
\label{sec:general_magnetic_geometry}
Lurking within this theory was an assumption of straight field-line geometry which considerably simplified the derivations involved. It is clear that this assumption will generally not hold in any plasma of physical interest so let us address this point. 

We begin in general magnetic geometry, although we will later find it necessary to discuss trapped and circulating particles. The kinetic equation becomes
\begin{equation}
\pder[f]{t} + v_{\parallel}\pder[f]{l} - \pder{\mu}\left(\frac{\mu}{\tau_{r}}f\right) - \frac{\mu}{m}\nabla_{\parallel}B \pder[f]{v_{\parallel}} =  {C}_{ee} + {C}_{ei} 
\end{equation}
where $l$ parametrises the length of a magnetic field line, and we retain the assumption that the parallel electric field is negligible. Here we can exploit the existence of two timescales by expanding in the small parameter
\begin{equation}
\epsilon_{1} \equiv \frac{\tau_{b}}{\tau_{r}} \ll 1
\end{equation}
where $\tau_{b} \sim L/v$ denotes the bounce time i.e., the typical time is takes to travel the macroscopic distance $L$ along the field. Thus, writing $f = f_{0} + f_{1} + \cdots,$ one obtains at leading order 
\begin{equation}
v_{\parallel} \pder[f_{0}]{l} - \frac{\mu}{m}\nabla_{\parallel}B \pder[f_{0}]{v_{\parallel}} = 0.
\label{zeroeth_order_kinetic equation}
\end{equation} 
One can solve this equation by making the change of variables $(l,\mu,v_{\parallel}) \rightarrow (l,\mu,w)$ where we have introduced the total particle energy\begin{equation}
w = \frac{mv_{\parallel}^{2}}{2} + \mu B,
\end{equation}  so that the kinetic equation becomes at leading order
\begin{equation}
v_{\parallel}\left(\pder[f_{0}]{l} + \mu \nabla_{\parallel}B \pder[f_{0}]{w}\right)- \frac{\mu}{m}\nabla_{\parallel}B \left(mv_{\parallel}\pder[f_{0}]{w}\right) = v_{\parallel}\left(\pder[f_{0}]{l}\right)_{w} = 0,
\end{equation}
which has the solution for the lowest order distribution function
\begin{equation}
f_{0} = f_{0}(\mu,w).
\end{equation}

At the next order, the kinetic equation becomes
\begin{equation}
\pder[f_{0}]{t} - \frac{1}{\tau_{r}} \pder{\mu}\left(\mu f\right) - \mu\frac{B}{\tau_{r}}\pder[f]{w}  +  v_{\parallel}\left(\pder[f_{1}]{l}\right)_{w} = C_{ee}(f_{0}) + C_{ei}(f_{0}).
\label{first_order_kinetic_equation} 
\end{equation}

\subsection{The bounce-averaged kinetic equation}

In order to remove the dependence on $f_{1},$ we define the bounce-average of a function $Q(\mu,w,l)$ by
\begin{equation}
\overline{Q}(\mu,w) = \int Q(\mu,w,l) \frac{\dd l}{v_{\parallel}} \bigg/ \int \frac{\dd l}{v_{\parallel}},
\label{bounceavrg}
\end{equation}
with
\begin{equation}
v_{\parallel} = \sqrt{\frac{2}{m}(w-\mu B)}.
\end{equation}
The integral in (\ref{bounceavrg}) is taken between consecutive bounce points, defined by $v_{\parallel} = 0,$ for trapped particles. For circulating particles, the integral is taken once around a field line if the field line is closed. If the field line is not closed, but instead traces out a magnetic surface, as in a stellarator or tokamak, then the bounce-average for circulating particles is given by
\begin{equation}
\overline{Q}(\mu,w) =\lim\limits_{L \rightarrow \infty} \int_{-L}^{L} Q(\mu,w,l) \frac{\dd l}{v_{\parallel}} \bigg/ \int_{-L}^{L} \frac{\dd l}{v_{\parallel}}.
\end{equation}
Taking the bounce average of (\ref{first_order_kinetic_equation}) then yields 
\begin{equation}
\pder[f_{0}]{t} - \overline{\frac{1}{\tau_{r}}\pder{\mu}(\mu f_{0})} - \mu \overline{\left(\frac{B}{\tau_{r}}\right)}\pder[f_{0}]{w} = \overline{C_{ee}(f_{0})} + \overline{C_{ei}(f_{0})}.
\end{equation}
Now, here we must be careful when bounce-averaging as $\tau_{r} \propto B^{-2}.$ Let us write
\begin{equation}
\tau_{r} = \tau_{0}\left(\frac{B}{B_{0}}\right)^{-2},
\end{equation}
where $B_{0}$ and $\tau_{0}$ appearing on the right-hand side of this equation are constants. We thus arrive at the bounce averaged kinetic equation
\begin{equation}
\pder[f_{0}]{t} - \frac{1}{\tau_{0}} \overline{\frac{B^{2}}{B_{0}^{2}}}\pder{\mu}\left(\mu f_{0}\right) - \frac{\mu}{\tau_{0}} \overline{\frac{B^{3}}{B_{0}^{2}}} \pder[f_{0}]{w} = \overline{C_{ee}(f_{0})} + \overline{C_{ei}(f_{0})}.
\label{bae}
\end{equation}
Since $\overline{B^{2}}$ and $\overline{B^{3}}$ are, in general, complicated functions of $\mu/w,$ one cannot hope to solve this equation in full generality. However, we can deduce the distribution function on a significantly long timescale $\tau$ which is larger than the radiation time and comparable to the collision time $\tau_{r} \ll \tau \lesssim \tau_{c}$. In this instance, the perpendicular energy will have been radiated away to leading order and we have 
\begin{equation}
\lambda B \ll 1,
\end{equation}
where we have introduced $\lambda = \mu/w.$ Trapped particles will thus be absent from the population. In this regime, one can neglect the third term in equation (\ref{bae}) and thus obtain 
\begin{equation}
\pder[f_{0}]{t} - \frac{1}{\tau_{0}}\overline{\frac{B^{2}}{B_{0}^{2}}}\pder{\mu}\left(\mu f_{0}\right) = \overline{C_{ee}(f_{0})} + \overline{C_{ei}(f_{0})}.
\label{kinetic_equation_curved}
\end{equation}
Moreover, we can also evaluate 
\begin{equation}
\overline{\frac{B^{2}}{B_{0}^{2}}} = \frac{1}{B_{0}^{2}}\oint \frac{B^{2}(l)}{\sqrt{1-\lambda B}} \dd l \bigg/ \oint\frac{\dd l}{\sqrt{1-\lambda B}} = \frac{1}{B_{0}^{2}}\oint B^{2} \dd l \bigg/ \oint \dd l \equiv 1
\end{equation}
where the last equivalence follows from the choice to simply define $B_{0}^{2}$ to be the average of $B^{2}$ over a field line. 

Hence, we arrive at 
\begin{equation}
\pder[f_{0}]{t} - \frac{1}{\tau_{0}}\pder{\mu}(\mu f_{0})  = \overline{C_{ee}(f_{0})} + \overline{C_{ei}(f_{0})}.
\label{bae2}
\end{equation}

\subsection{The bounce-averaged collision operator}

We have already shown that at leading order, the contribution from electron-ion collisions is formally smaller than electron-electron collisions provided $n_{i}Z^{2}/n_{e} = O(1).$ Let us thus restrict our attention to the electron-electron collision operator. 

We know that at leading order
\begin{equation}
C_{ee}(f_{0}) \simeq \frac{\sigma g(\vp)}{\vpp}\pder{\vpp}\vpp \pder[f_{0}]{\vpp}.
\end{equation}
We must be cautious as $\vpp$ varies along the orbit and should not be used as a coordinate in the bounce-averaged equation. When the collision operator is instead written using the magnetic moment $\mu$ as a coordinate we obtain 
\begin{equation}
\overline{C_{ee}(f_{0})} \simeq 2\sigma m\overline{\frac{g(\vp)}{B}}  \pder{\mu}\left(\mu\pder[f_{0}]{\mu}\right)
\label{bouncearvgcollisions} 
\end{equation}
and thus
\begin{equation}
\pder[f_{0}]{t} - \frac{1}{\tau_{0}}\pder{\mu}(\mu f_{0}) = 2 \sigma m \overline{\frac{g(\vp)}{B}} \pder{\mu}\left(\mu \pder[f_{0}]{\mu}\right).
\end{equation}
It is of course no accident that this equation looks remarkably similar to the equation obtained in the straight field line limit. One can make the analogy exact and state that the leading order dynamics of both systems are governed by the differential equation provided we associate the $B^{2}$ appearing in Larmor's formula, and the function $g(\vp)$ arising from the electron-electron collision operator, with their averages over a magnetic field line. 

That is, on timescales $\tau$ with 
\begin{equation}
\tau_{b} \ll \tau_{r} \ll \tau \lesssim \tau_{c}
\end{equation}
the distribution function will become strongly anisotropic in general magnetic geometry.

\section{Applicability of this theory}
\label{sec:applicability_of_this_theory}

The theory developed in the preceding section sections applies to any optically thin plasma where the collision time exceeds the radiation emission time, which is always true if the density is sufficiently small. Two broad classes of applications can be envisaged. 

\subsection{Cyclotron sources in the laboratory}

Non-neutral plasmas, specifically plasmas consisting of charged particles with a single sign of charge can be confined in Penning-Malmberg traps \citep{Dubin1999a}, \citep{Danielson2015a}. Such a trap consists of a vacuum region inside an electrode structure consisting of a stack of hollow, metal cylinders. A uniform axial magnetic field is then applied to inhibit particle motion in the radial direction. Voltages must also be  imposed on the end electrodes to prevent particle loss in the magnetic field direction.

 It is frequently useful to compress plasmas radially; for instance, to increase the plasma density. This usually accomplished by applying a torque on the plasma using rotating electric fields, the so-called ``rotating wall technique'' \citep{Anderegg1998a}. Very long confinement times(on the order of hour or days) can be achieved using these techniques, making their use highly desirable. Particle cooling is often necessary to maintain good confinement by mitigating the heating caused by the torque using the rotating wall method. In the case of electrons or positrons, if the magnetic field is sufficiently strong, the particles will cool by cyclotron radiation \citet{ONeil1985}. An example of where cyclotron cooling is employed successfully to this end is in the production of anti-hydrogen, where this process is used cool pure electron plasma to sub eV temperatures \citet{Amoretti2002}.

Another major application of the theory developed in this work will be for the first laboratory experiment, currently under development, to create and confine the first terrestrial electron-positron plasmas in the laboratory. This is done by first accumulating positrons from a powerful source and then injecting these into a pure electron plasma confined by the dipolar magnetic field of a current-carrying circular coil, so that a stationary, quasineutral electron-positron plasma is formed \citep{SunnPedersen2012}. This system will satisfy the necessary conditions of being optically thin and with a radiation time which should be initially shorter than the collision time. However, there is an additional complication in that the plasma will be so strongly magnetized that Coulomb collisions are no longer effectively described by the Landau operator. The theory of scattering in strongly-magnetized anisotropic pair plasmas is developed in the companion paper (II).

\subsection{Synchrotron sources in astrophysics}

This work was built on the fact that a charged particle moving in a magnetic field radiates energy. At non-relativistic energies, the focus of this paper, this process is called cyclotron-cooling. At relativistic velocities it is known as synchrotron radiation.

 Synchrotron sources are ubiquitous and the emission of relativistic and ultra-relativistic electrons gyrating in a magnetic field is a process which dominates much of high energy astrophysics. Indeed, it is known that synchrotron radiation is responsible for the non-thermal optical and
X-ray emission observed in the Crab Nebula \citep{Rees_Crab} and that pulsars are strong synchrotron sources  \citep{Sturrock_Pulsars}. 

Of course, the physics of these systems should also be treated carefully using a model taking relativistic effects into account in the modelling of the radiation term and the collision terms. This is an area of active research and a potential axis along which this current work could be further developed. There also likely exist analogous systems in astrophysics where the electrons are sub-relativistic but nevertheless are still strongly radiating and weakly collisional in the sense discussed here. Such a system would of course still be required to satisfy \begin{equation}
		\frac{\tau_{r}}{\tau_{c}} =  \left(\frac{\omega_{p}}{\Omega_{c}}\right)^{2}\left(\frac{c}{v_{\text{th}}}\right)^{3} \ll 1.
\end{equation}
The easiest way to satisfy this is of course to allow relativistic plasmas, but we might also envision some exotic astrophysical plasma which is sufficiently nebulous (very low density) but also sufficiently strongly magnetised (very strong magnetic fields) as to allow this condition to be met in a non-relativistic plasma.  In such systems, the theory in this paper will be directly relevant.

However, these is another caveat here which must be carefully considered. We have presented here a mechanism through which plasmas can become strongly anisotropic in velocity space. In actuality, there is a plethora of instabilities which could act on astrophysical plasmas and restore isotropy on timescales shorter than the collision time. It seems to be that the question of (in)stability could depend sensitively on plasma parameters and that considerations of isotropy-restoring instabilities do not necessarily preclude the types of systems studied in this work. 

\section{Conclusions}

In this paper, we have developed a theory for collisional scattering in strongly anisotropic plasmas. Such plasmas arise due to the emission of radiation when charged particles move in magnetic fields, which leads to rapid depletion of the perpendicular kinetic energy. We have derived equations which describe the evolution of the electron distribution function in such plasmas. Unusually for plasma physics, the collision operator could be calculated analytically, albeit only to logarithmic accuracy, although the distribution function is far from Maxwellian.

It was found that in such strongly-anisotropic populations, the evolution of the lowest order distribution function is dominated by electron-electron collisions unless $n_{i}Z^{2}/n_{e} \gg 1.$ Such collisions lead to a distribution that is Maxwellian in $\vpp$ for any value of $\vp.$  

The distribution over $\vp$ can be ascertained from an equation governing the density (in $\vp$ space) $N,$ the integral of the distribution function over perpendicular velocities. We found that that this quantity satisfies an ill-posed, non-linear, parabolic PDE, reminiscent of the backwards diffusion equation. 

This equation can be solved in two limits: firstly in the case where the contribution from electron-ion collisions can be neglected, in which case a similarity solution is found; and secondly, in the case where the contribution from electron-electron collisions can be neglected and the equation can be solved by integration along the characteristic curves. In this latter limit, the equation is well posed. In both cases, the remaining energy is radiated on the time scale of the ordinary collision frequency divided by $|\ln \epsilon|.$ 

Several candidates for areas of application of this theory were presented. These included both astrophysical applications and experimental applications. The latter of these applications will be developed in a companion paper (II). 

\newpage
\bibliographystyle{abbrvnat}
\bibliography{radiative_cooling_1.bib}

\begin{thebibliography}{14}
\providecommand{\natexlab}[1]{#1}
\providecommand{\url}[1]{\texttt{#1}}
\expandafter\ifx\csname urlstyle\endcsname\relax
  \providecommand{\doi}[1]{doi: #1}\else
  \providecommand{\doi}{doi: \begingroup \urlstyle{rm}\Url}\fi

\bibitem[Amoretti et~al.(2002)Amoretti, Carraro, Lagomarsino, Manuzio, Testera,
  and Variola]{Amoretti2002}
M.~Amoretti, C.~Carraro, V.~Lagomarsino, G.~Manuzio, G.~Testera, and
  A.~Variola.
\newblock {Electron plasma for antiproton cooling in the ATHENA experiment}.
\newblock \emph{AIP Conference Proceedings}, 606:\penalty0 45, 2002.
\newblock \doi{10.1063/1.1454266}.
\newblock URL \url{https://doi.org/10.1063/1.1454266}.

\bibitem[Anderegg et~al.(1998)Anderegg, Hollmann, and Driscoll]{Anderegg1998a}
F.~Anderegg, E.~M. Hollmann, and C.~F. Driscoll.
\newblock {Rotating field confinement of pure electron plasmas using
  trivelpiece-gould modes}.
\newblock \emph{Physical Review Letters}, 81\penalty0 (22):\penalty0
  4875--4878, 1998.
\newblock ISSN 10797114.
\newblock \doi{10.1103/PhysRevLett.81.4875}.

\bibitem[Andersson et~al.(2001)Andersson, Helander, and
  Eriksson]{Andersson2001}
F.~Andersson, P.~Helander, and L.-G. Eriksson.
\newblock {Damping of relativistic electron beams by synchrotron radiation}.
\newblock \emph{Physics of Plasmas}, 8\penalty0 (12):\penalty0 5221--5229, dec
  2001.
\newblock ISSN 1070-664X.
\newblock \doi{10.1063/1.1418242}.
\newblock URL \url{http://aip.scitation.org/doi/10.1063/1.1418242}.

\bibitem[Braginskii(1965)]{Braginskii1965}
S.~I. Braginskii.
\newblock {Transport Processes in a Plasma}.
\newblock \emph{RvPP}, 1:\penalty0 205, 1965.

\bibitem[Danielson et~al.(2015)Danielson, Dubin, Greaves, and
  Surko]{Danielson2015a}
J.~R. Danielson, D.~H. Dubin, R.~G. Greaves, and C.~M. Surko.
\newblock {Plasma and trap-based techniques for science with positrons}.
\newblock \emph{Reviews of Modern Physics}, 87\penalty0 (1):\penalty0 247--306,
  mar 2015.
\newblock ISSN 15390756.
\newblock \doi{10.1103/RevModPhys.87.247}.

\bibitem[Dubin and O'Neil(1999)]{Dubin1999a}
D.~H. Dubin and T.~M. O'Neil.
\newblock {Trapped nonneutral plasmas, liquids, and crystals (the thermal
  equilibrium states)}.
\newblock \emph{Reviews of Modern Physics}, 71\penalty0 (1):\penalty0 87--172,
  jan 1999.
\newblock ISSN 00346861.
\newblock \doi{10.1103/revmodphys.71.87}.

\bibitem[Hazeltine and Mahajan(2004)]{Hazeltine_Mahajan}
R.~D. Hazeltine and S.~M. Mahajan.
\newblock {Radiation reaction in fusion plasmas}.
\newblock \emph{Physical Review E - Statistical Physics, Plasmas, Fluids, and
  Related Interdisciplinary Topics}, 70\penalty0 (4):\penalty0 6, oct 2004.
\newblock ISSN 1063651X.
\newblock \doi{10.1103/PhysRevE.70.046407}.

\bibitem[Krall and Trivelpiece(1986)]{Krall1986}
N.~A. Krall and A.~W. Trivelpiece.
\newblock \emph{{Principles Of Plasma Physics}}.
\newblock San Francisco Press, 1986.

\bibitem[Landau(1936)]{LDLandau1936}
L.~D. Landau.
\newblock {Die kinetische Gleichung fuer den fall Coulombscher wechselwirkung}.
\newblock \emph{Sowjet, Phys. Z.}, 1936.

\bibitem[{Low}(1973)]{Low}
B.~C. {Low}.
\newblock {Resistive Diffusion of Force-Free Magnetic Fields in a Passive
  Medium. 11. a Nonlinear Analysis of the One-Dimensional Case}.
\newblock \emph{The Astrophysical Journal}, 184:\penalty0 917--930, Sept. 1973.
\newblock \doi{10.1086/152382}.

\bibitem[O'Neil(1985)]{ONeil1985}
T.~M. O'Neil.
\newblock {New Theory of Transport Due to Like-Particle Collisions}.
\newblock \emph{Physical Review Letters}, 55\penalty0 (9):\penalty0 943--946,
  aug 1985.
\newblock ISSN 0031-9007.
\newblock \doi{10.1103/PhysRevLett.55.943}.
\newblock URL \url{https://link.aps.org/doi/10.1103/PhysRevLett.55.943}.

\bibitem[Rees and Gunn(1974)]{Rees_Crab}
M.~J. Rees and J.~E. Gunn.
\newblock {The Origin of the Magnetic Field and Relativistic Particles in the
  Crab Nebula}.
\newblock \emph{Monthly Notices of the Royal Astronomical Society},
  167\penalty0 (1):\penalty0 1--12, apr 1974.
\newblock ISSN 0035-8711.
\newblock \doi{10.1093/mnras/167.1.1}.

\bibitem[Sturrock(1971)]{Sturrock_Pulsars}
P.~A. Sturrock.
\newblock {A MODEL OF PULSARS}.
\newblock \emph{The Astrophysical Journal}, 164:\penalty0 529--556, 1971.

\bibitem[{Sunn Pedersen} et~al.(2012){Sunn Pedersen}, Danielson, Hugenschmidt,
  Marx, Sarasola, Schauer, Schweikhard, Surko, and Winkler]{SunnPedersen2012}
T.~{Sunn Pedersen}, J.~R. Danielson, C.~Hugenschmidt, G.~Marx, X.~Sarasola,
  F.~Schauer, L.~Schweikhard, C.~M. Surko, and E.~Winkler.
\newblock {Plans for the creation and studies of electron–positron plasmas in
  a stellarator}.
\newblock \emph{New Journal of Physics}, 14\penalty0 (3):\penalty0 035010, mar
  2012.
\newblock ISSN 1367-2630.
\newblock \doi{10.1088/1367-2630/14/3/035010}.
\newblock URL
  \url{http://stacks.iop.org/1367-2630/14/i=3/a=035010?key=crossref.5b2a358e6b2ab47a4970878e038cfd8b}.

\end{thebibliography}

\appendix
\section{Moments of the collision operator}
\label{app1}

Let us now evaluate the contributions to (\ref{a}) arising from the moments of the collision operator. 
\subsection{Calcuation of \texorpdfstring{$I_{ee}$}{TEXT}}

The contribution to (\ref{a}) arising from electron-electron collisions is given by 
\begin{align}
	&\int_{0}^{\infty} C_{ee}(f) \, \dd^{2}\bm{v}_{\pp} = \sigma\pder{\vp} \int_{0}^{\infty} \dd^{2}\bm{v}_{\perp} \int ff^{\prime} \frac{u^{2}\bm{\mathrm{b}} - u_{\pa}\bm{u}}{u^{3}} \cdot \left(\graddv \ln f - \graddv^{\prime} \ln f^{\prime} \right) \, \dd^{3}\bm{v}^{\prime}, \nonumber \\
	&= \sigma \pder{\vp}\int_{0}^{\infty} \dd^{2}\bm{v}_{\pp} \int ff^{\prime} \left[\frac{u_{\pp}^{2}}{u^{3}}\left(\pder[\ln f]{\vp} - \pder[\ln f^{\prime}]{\vp^{\prime}}\right) - \frac{u_{\pa}\bm{u}_{\perp}}{u^{3}} \cdot \left(\graddv \ln f  - \graddv^{\prime} \ln f^{\prime}\right)\right] \, \dd^{3}\bm{v}^{\prime}, \nonumber \\ &= \sigma \pder{\vp} \int_{0}^{\infty}f\, \dd^{2}\bm{v}_{\perp} \int \frac{f^{\prime}}{u^{3}}\left[u_{\pp}^{2}\left(\pder[\ln f]{\vp} - \pder[\ln f^{\prime}]{\vp^{\prime}}\right) + u_{\pa}\bm{u}_{\perp} \cdot \left(\frac{m\bm{v}_{\pp}}{T_{\pp}} - \frac{m\bm{v}_{\pp}^{\prime}}{T_{\pp}^{\prime}}\right)\right] \, \dd^{3}\bm{v}^{\prime},
\end{align}
where $T_{\pp}^{\prime} = T_{\pp}(\vp^{\prime}).$ 

In order to carry out the integrals over $\bm{v}_{\pp}$ and $\bm{v}_{\pp}^{\prime}$ we note that:
\begin{gather}
	u_{\perp}^{2} = \vpp^{2} + \vpp^{\prime 2} - 2\vpp\vpp^{\prime}\cos\theta, \\
	\bm{u}_{\pp}\cdot \bm{v}_{\pp} = \vpp^{2} - \vpp\vpp^{\prime}\cos\theta, \\
	\bm{u}_{\pp}\cdot\bm{v}_{\pp}^{\prime} = -\vpp^{\prime 2} + \vpp\vpp^{\prime}\cos\theta,
\end{gather}
where $\theta$ is the angle between $\bm{v}_{\pp}$ and $\bm{v}_{\pp}^{\prime}.$ We note further that upon integration, any terms involving $\cos\theta$ will vanish. Hence, we may write
\begin{align}
	\int_{0}^{\infty}C_{ee}(f) \, \dd^{2}\bm{v}_{\perp} = \sigma \pder{\vp} \int_{0}^{\infty} \dd^{2}\bm{v}_{\pp} \int \frac{1}{u^{3}} \left[(\vpp^{2} + \vpp^{\prime 2}) \left(f^{\prime}\pder[f]{\vp} - f \pder[f^{\prime}]{\vp^{\prime}}\right)\right.& \nonumber \\  \left.+u_{\pa}ff^{\prime}\left(\frac{m\vpp^{2}}{T_{\perp}} + \frac{m\vpp^{\prime 2}}{T_{\pp}^{\prime}}\right)\right.&\left. \vphantom{\frac{1}{2}}\right] \, \dd^{3}\bm{v}^{\prime}.
	\label{411}
\end{align}
The leading order contribution to this integral comes from the final term in equation(\ref{411}) and so we have 
\begin{equation}
	\int_{0}^{\infty} C_{ee}(f) \, \dd^{2}\bm{v}_{\pp} \simeq \sigma \pder{\vp} \int \dd^{2}\bm{v}_{\pp}\int \frac{u_{\pa}}{u^{3}}ff^{\prime}\left(\frac{m\vpp^{2}}{T_{\pp}} + \frac{m\vpp^{\prime 2}}{T_{\pp}^{\prime}}\right) \dd^{3} \bm{v}^{\prime}.
\end{equation}

We perform the $\vp^{\prime}-$integral first and are thus led to consider 
\begin{equation}
	\int_{-\infty}^{\infty} \frac{u_{\pa}}{u^{3}}f^{\prime}\, \dd v_{\pa}^{\prime} = \int_{-\infty}^{\infty}  \frac{m N(\vp^{\prime})}{2\pi T_{\pp}(\vp^{\prime})}\exp \left(\frac{m\vpp^{\prime 2}}{2T_{\pp}(\vp^{\prime})}\right)\frac{\vp - \vp^{\prime}}{\left((\vp - \vp^{\prime})^{2} + |\bm{v}_{\pp} - \bm{v}_{\pp}^{\prime}|^{2}\right)^{3/2}} \, \dd \vp^{\prime},
\end{equation}
which is an integral of the form 
\begin{equation}
	\int_{-\infty}^{\infty} \frac{x}{(x^{2} + \epsilon^{2})^{3/2}} f(x) \, \dd x = \int_{-\infty}^{\infty} \frac{\dd f}{\dd x}\frac{1}{\sqrt{x^{2} + \epsilon^{2}}} \, \dd x = 2\frac{\dd f}{\dd x} \bigg\vert_{x = 0} (| \ln \epsilon| + O(1)), 
\end{equation}
and thus becomes
\begin{equation}
	\int_{-\infty}^{\infty} \frac{u_{\pa}}{u^{3}}f^{\prime} \, \dd \vp^{\prime} = - 2\pder[f^{\prime}]{\vp^{\prime}} \bigg\vert_{\vp^{\prime} = \vp} (| \ln \epsilon| + O(1)). 
\end{equation}
Hence, we arrive at 
\begin{align}
	I_{ee} \equiv \int_{0}^{\infty} C_{ee}(f) \, \dd^{2}\bm{v}_{\perp} &= - 2\sigma |\ln \epsilon| \pder{\vp} \int f \, \dd^{2}\bm{v}_{\perp} \int \pder[f^{\prime}]{\vp^{\prime}} \bigg\vert_{\vp^{\prime} = \vp} \, \dd^{2}\bm{v}_{\perp}^{\prime}, \nonumber \\ &= -2\sigma |\ln \epsilon| \pder{\vp} \left(N \pder[N]{\vp}\right), \nonumber \\ &= -\sigma |\ln \epsilon| \pder[^{2}N^{2}]{\vp^{2}}.
\end{align}

\subsection{Calculation of \texorpdfstring{$I_{ei}$}{TEXT}}

The contribution arising from electron-ion collisions gives 
\begin{align}
I_{ei} \equiv	\int C_{ei}(f) \, \dd^{2}\bm{v}_{\pp} &= Z\sigma \pder{\vp} \int \bm{\mathrm{b}} \cdot \left(\frac{v^{2}\mathsfbi{I} - \bm{v}\bm{v}}{v^{3}} \cdot \graddv f \right) \, \dd^{2} \bm{v}_{\perp},  \nonumber \\ &= Z\sigma \pder{\vp} \int \left(\frac{\vpp^{2}}{v^{3}}\pder[f]{\vp} - \frac{\vp}{v^{3}}\bm{v}_{\perp}\cdot \graddv f\right) \, \dd^{2}\bm{v}_{\perp}, \nonumber \\
	&\simeq Z\sigma \pder{\vp} \int \frac{2f}{\vp |\vp|} \, \dd^{2}\bm{v}_{\perp}, \nonumber 
	\\ &= Z\sigma \pder{\vp}\left(\frac{N}{\vp|\vp|}\right).
\end{align}
We remark here that, although the electron-ion collision operator was ordered much smaller than the electron-electron collision operator, it is not necessary for $I_{ei}$ to be ordered much smaller than $I_{ee}.$ 

\section{Asymptotic forms of \texorpdfstring{$N$}{TEXT}}
\label{appB}
In this appendix, we calculate the long-time $(y \gg 1)$ and short-time $(y\ll 1)$ asymptotic behaviour of (\ref{512}).
\subsection{Long-time limit}
\label{appB_long}
For $y\gg1,$ most of the contribution to this integral comes from $x^{3} \ll y,$ so that 
\begin{align}
	\frac{W_{\pp}(y)}{W_{\pp}(0)} &= \frac{2}{\sqrt{\pi}} \int_{0}^{\infty} \left(\frac{x}{y^{1/3}}\right)^{4} \exp \left[-y^{2/3}\left(\frac{x^{3}}{y} + 1\right)^{2/3} \right] \, \dd x, \nonumber \\ &\simeq \frac{2}{\sqrt{\pi}}\int_{0}^{\infty} \left(\frac{x}{y^{1/3}}\right)^{4} \exp \left[-y^{2/3}\left(1 + \frac{2x^{3}}{3y}\right)\right]\, \dd x.
\end{align}  
We now make the substitution $s = 2x^{3}/3y^{1/3}$ to obtain 
\begin{align}
	\frac{W_{\pp}(y)}{W_{\pp}(0)} &\simeq \frac{2}{\sqrt{\pi}} \int_{0}^{\infty} \frac{1}{y^{4/3}} \left(\frac{3y^{1/3}}{2}\right)^{4/3} s^{4/3} \exp(-y^{2/3} - s) \frac{\dd s}{3s^{2/3}}.
\end{align}
Thus, recalling the definition of the Gamma function,
\begin{equation}
	\Gamma(z) = \int_{0}^{\infty} x^{z-1} \exp (-x) \, \dd x,
\end{equation}
we arrive at 
\begin{equation}
	\frac{W_{\pp}(y)}{W_{\pp}(0)} \simeq \frac{1}{\sqrt{\pi}}\left(\frac{3}{2}\right)^{2/3} \Gamma \left(\frac{5}{3}\right) y^{-7/9} \exp( -y^{2/3}), \quad\quad y \gg 1.
	\label{515}
\end{equation}

\subsection{Short-time limit}
\label{appB_short}
In the short-time limit, $y \ll 1$ we have 
\begin{align}
	\frac{\dd}{\dd y} \left(\frac{W_{\pp}(y)}{W_{\pp}(0)}\right) &= -\frac{2}{\sqrt{\pi}} \int_{0}^{\infty} x^{4} \exp(-(x^{3} + y)^{2/3})\left(\frac{4}{3(x^{3} + y)^{7/3}} + \frac{2}{3(x^{3} + y)^{5/3}}\right) \, \dd x, \nonumber \\ 
	&\simeq -\frac{8}{3\sqrt{\pi}} \int_{0}^{\infty} \frac{x^{4}}{(x^{3}+y)^{7/3}} \exp (-(x^{3} + y)) \, \dd x,
\end{align}
where most of the contribution comes from the region $x \sim y^{1/3}$ where the integrand is of order 
\begin{equation}
	\frac{x^{4}}{(x^{3}+y)^{7/3}} \exp (-(x^{3} + y)) \sim \frac{y^{4/3}}{y^{7/3}} = \frac{1}{y},
\end{equation}
and we thus expect the integral to be of order $y^{-2/3}.$ Indeed, 
\begin{equation}
	\frac{\dd}{\dd y} \left(\frac{W_{\pp}(y)}{W_{\pp}(0)}\right) \simeq -\frac{8}{3\sqrt{\pi}} \int_{0}^{\infty} \frac{x^{4}}{(x^{3} + y)^{7/3}} \, \dd x = -\frac{8}{9\sqrt{\pi}} \int_{0}^{\infty} \frac{u^{2/3}}{(u+y)^{7/3}} \dd u,
\end{equation}
where we have made the substitution $u = x^{3}.$ Upon making a further substitution $p = u/y,$ we obtain 
\begin{equation}
	\frac{\dd}{\dd y} \left(\frac{W_{\pp}(y)}{W_{\pp}(0)}\right) \simeq -\frac{8}{9\sqrt{\pi} y^{2/3}}\int_{0}^{\infty} \frac{p^{2/3}}{(1+p)^{7/3}} \, \dd p,
\end{equation} 
where we recognise 
\begin{equation}
	\int_{0}^{\infty} \frac{p^{\alpha}}{(1+p)^{\beta}} \,\dd p = \frac{\Gamma(\alpha +1)\Gamma(\beta - \alpha - 1)}{\Gamma(\beta)}.
\end{equation}
Thus, we may conclude that 
\begin{equation}
	\frac{\dd }{\dd y} \left(\frac{W_{\pp}(y)}{W_{\pp}(0)}\right) \simeq -\frac{8}{9\sqrt{\pi}y^{2/3}} \frac{\Gamma(5/3) \Gamma(2/3)}{\Gamma(7/3)},
\end{equation}
which leads to 
\begin{equation}
	\frac{W_{\pp}(y)}{W_{\pp}(0)} \simeq 1 - \frac{8}{3\sqrt{\pi}}\frac{\Gamma(5/3) \Gamma(2/3)}{\Gamma(7/3)} y^{1/3}, \quad\quad y \ll 1. \label{521}
\end{equation}
This equation can be rewritten in terms of the time $t$ and the Braginskii electron-ion collision time $\tau_{ei}$ as
\begin{equation}
	\frac{W_{\pp}(t)}{W_{\pp}(0)} \simeq 1 - \frac{\Gamma(5/3) \Gamma(2/3)}{\Gamma(7/3)} \left(\frac{128}{3\pi}\right)^{1/6} \left(\frac{t}{\tau_{ei}}\right)^{1/3}. \label{522}
\end{equation}

\end{document}